# Moving dunes on the Google Earth

**Amelia Carolina Sparavigna**
**Department of Applied Science and Technology, Politecnico di Torino, Italy**

*Several methods exist for surveying the dunes and estimate their migration rate. Among methods suitable for the macroscopic scale, the use of the satellite images available on Google Earth is a convenient resource, in particular because of its time series. Some examples of the use of this feature of Google Earth are here proposed.*

**Keywords**: Dunes, Dune Migration, Satellite Imagery, Google Earth, Image Processing

In several regions of the world, the motion of sand dunes, though it seems quite slow, can become a challenge or even a threat for human activities. Some engineering solutions are available for protecting the sites, such as methods for surveying the dune fields and estimate their migration rate. The most common manner of preventing sand from advancing, is to create some fences across the path of the prevailing wind. These fences are trapping the movement of sand [1], because they slow down the flow of air, causing air to release its load. Experiments were implemented in sandy lands to evaluate the effects of different methods for stabilizing sand dune by means of vegetation restoration [2], creating an environment suitable for it [3]. However, in hyper-arid area, these methods are rather challenging [4].

Sand dunes are hills of sand with different forms and sizes, which can move. As complex systems of particles, the dunes and their motion have attracted several experimental studies in the past [5,6] and recently [7], such as many numerical simulations [8-12]. Let us remember that the first systematic study of the motion of dunes was due to Ralph Bagnold, who during 1940s did measurements directly on natural dunes, but also experiments in wind tunnels [13]. This methodology is still used for the collection of experimental data [14].

The researches on dune are important, because the knowledge of the dynamics of dunes can help in predicting their behavior, in particular when the dunes can directly threaten human activities. According to the United States Geological Survey, sand dunes are becoming more mobile as the climate changes [15]. Therefore, a monitoring of dune movement is fundamental and it can be accomplished through a number of methods, combining surface mapping with aerial and satellite imagery, GPS and LIDAR measurements. Then the monitoring of dune can range from the macro-scale of satellite data to the micro-scale of local sensors distributed on the surface of dunes. A recently proposed method, suitable for monitoring the consolidation of surfaces, is the 3D laser scanning [16].

For the large scale monitoring, historical archives containing aerial and satellite imagery can be useful to compile a database for each dune field. Some maps can be created, such as the one proposed in Reference 15, showing the evolution starting from 1953 of the profile of the field of dunes near the Grand Falls, Arizona.

In [17], it was discussed the use of satellite images freely available on the World Wide Web, as a convenient resource for the planning of future human settlements and activities, when compared with resources from archives and research journals. Here I am proposing the use of Google Earth and its time series of images to collect data on the motion of dunes. To this purpose we can simply use the time slider in Google Earth, to see the series of images, each reporting the day of recording. When local meteorological data were available, we could accompanying the images with these data, and then the migration of the dunes be linked to the environmental conditions, monitoring the actual changes. Here in the following some case studies are proposed.

1. **The dune field near the Grand Falls, Arizona.**

In the Reference 15, the dune field near the Grand Falls has been discussed. The researchers used archival imagery and plotted the migration of the dune field from 1953 to 2010. The dominant wind direction has been from the southwest throughout the time period studied. They concluded that these dunes are migrating at a rapid rate, more than 100 m in just 5 years.

Using Google Earth, after choosing the location of the Grand Falls, we have at our disposal the satellite images corresponding to years 1992, 2007, 2010 and 2012. To evaluate the speed of the dunes, in particular of the front of the field, we have to determine it in the image. For this reason, the image is processed using GIMP (the GNU Image Manipulation Program) [18], applying the filter for the edge detection with the Sobel method. In the following figure (Figure 1), we see the dune field of 2007 on the left, and on the right the image after the edge detection. With Paint software, the front of the field is marked. The same is repeated for the four images.

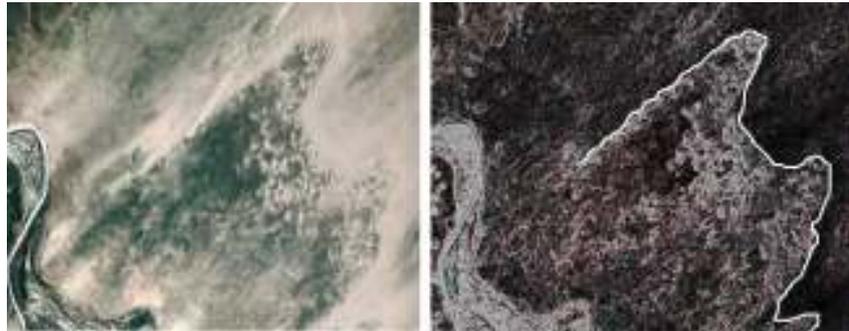

Figure 1 – On the left the dune field in the Google Earth images of 2007. On the right, the same image after processing with the edge detection of GIMP. The front is marked with a white line.

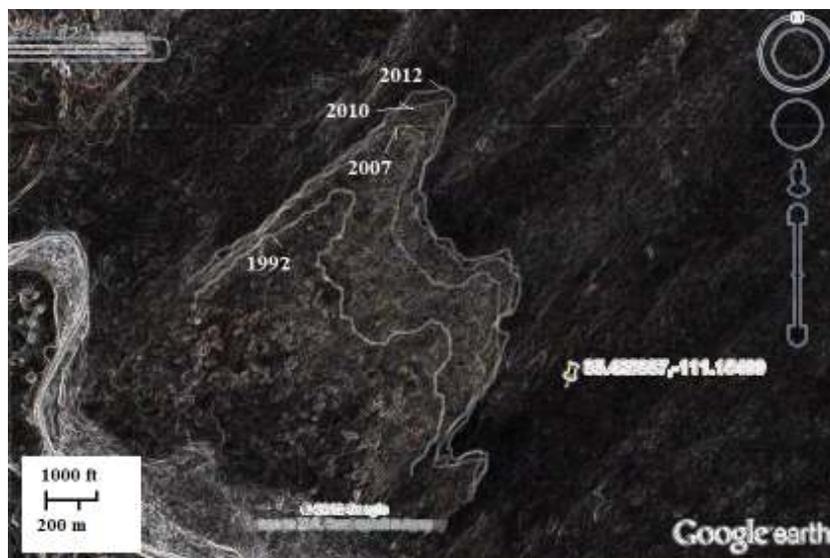

Figure 2 – Superposition of the images from Google Earth recorded at four different times (9/22/1992, 6/21/2007, 9/10/2010, 6/4/2012), after processing with GIMP.
The fronts of the field of dunes are labeled with years.

In the Figure 2, we see the superposition, made by GIMP, of the four filtered image corresponding to four different years. The speed of the front is, in some parts of it, in agreement with the speed given in Reference 15.

2. **Peruvian barchans.**

The dunes can have several shapes, according to the quantity of sand, the surface on which the sand moves and the blowing of the winds. These ridged hills have a windward side and a slip side in the lee of wind. When the ridge is arc-shaped, the dune is a "barchan". The barchan possesses then two "horns". We can observe isolated dunes or dune fields, where these dunes coalesced.

The coastal area of Peru is rich of dune fields and of isolated barchans. In Ref. 17 , I studied the dunes that fifty years ago  S. Parker Gay Jr., who at that time was a geophysicist on the exploration staff of Marcona Mining, studied. The Marcona mine is an huge open-pit iron mine, on the Pacific coast of Peru, 400 km southeast of Lima. In that period, a problem for this mining activity was the movement of large barchan sand dunes across the road that connected the mine with the shipping port of San Juan [17,19]. In [17] the Google Maps were compared with the maps of Ref.19.

Here we use just the images Google Earth. An interesting location is 15°08'19.60''S, 75°15'25.63''W. We have four images of June 10, 2009, August 23, 2009, June 10, 2011 and April 18, 2012. Each image is enhance by GIMP, as shown in the Fig.3; after, the fours images are superposed to have the map of Figure 4.

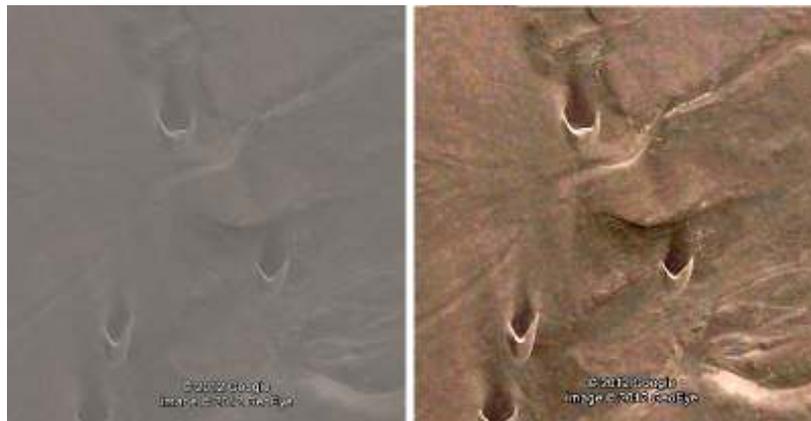

Figure 3 – Before combine the images, they need to be enhanced with GIMP. On the left the original image and on the right the image enhanced.

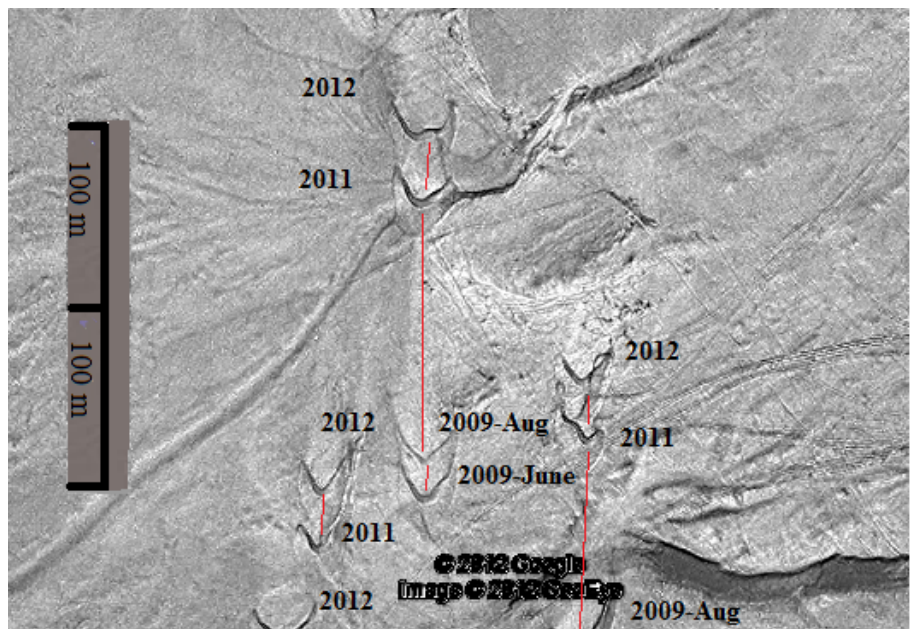

Figure 4 – The motion of the barchans is shown by the red lines connecting the positions of the dunes, each position labeled with years. The barchans move of 200 meters in three years.

After enhancing each image, sometimes it is better a conversion in grey tones; then combined together, we can obtain maps such as that of Figure 4. Using the scale we can measure the motion. Another interesting example of barchans and related motions can be obtained using the images at coordinates 15°06'48.03''S, 75°14'03.20''W. After processing, the result is shown in the Figure 5.

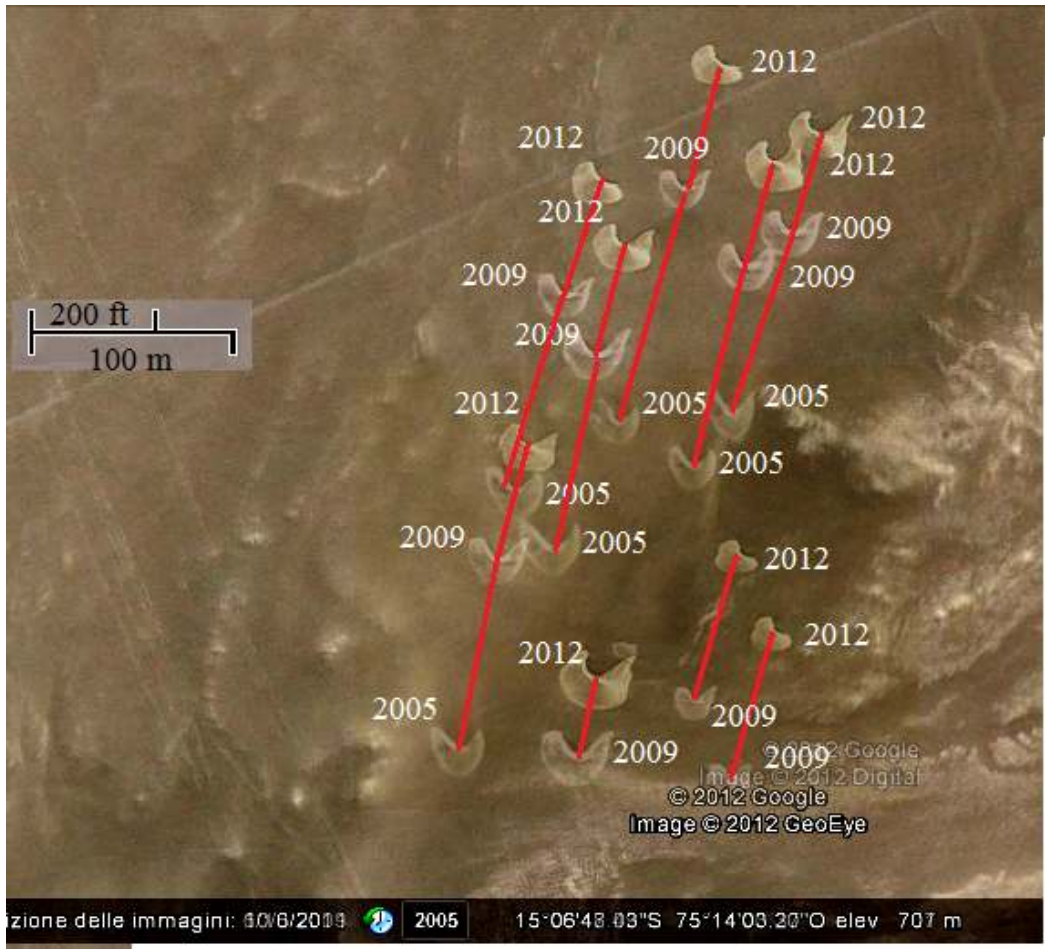

Figure 5 – Motion of barchans. The path of the dunes is given by the red lines, connecting the different positions labeled with years. Note that the small dunes move faster.

In the Figure 5 we can see that when a dune is small, it moves faster. In fact, the velocity of the dune depends on its height [20].
In [21], Wikipedia tells that several barchan dunes are near La Joya, Arequipa, Peru (16.714564S 71.834083W); "a number of dunes are readily visible from the Pan American Highway at the intersection with the Carretera Interoceanica just north of La Joya, where they can be seen passing over cement block buildings. In mid-2010, several smaller barchan dunes are approaching the Carretera from the south. These gray-colored dunes are formed from fine-grained volcanic ash from the Huaynaputina eruption in 1600." Therefore, we have an example of dunes moving towards human infrastructures. We can ask ourselves how fast are moving these dunes. An answer can come from Google Earth. Let us consider the dunes at coordinates 16°40'33.28''S, 71°50'20.09W: here there are five images in the time series: 4/23/2003, 6/11/2003, 2/25/2005, 11/20/2006 and 11/10/2009. Let us use two of them, those of 6/11/2003 and 11/10/2009. Then we see the motion over 6.4 years. After combining the two images we have the Figure 6. In the image, the reader can see a black line: since the two images are slightly shifted, I have marked a certain line on the ground in both images and used it to have a precise superposition.

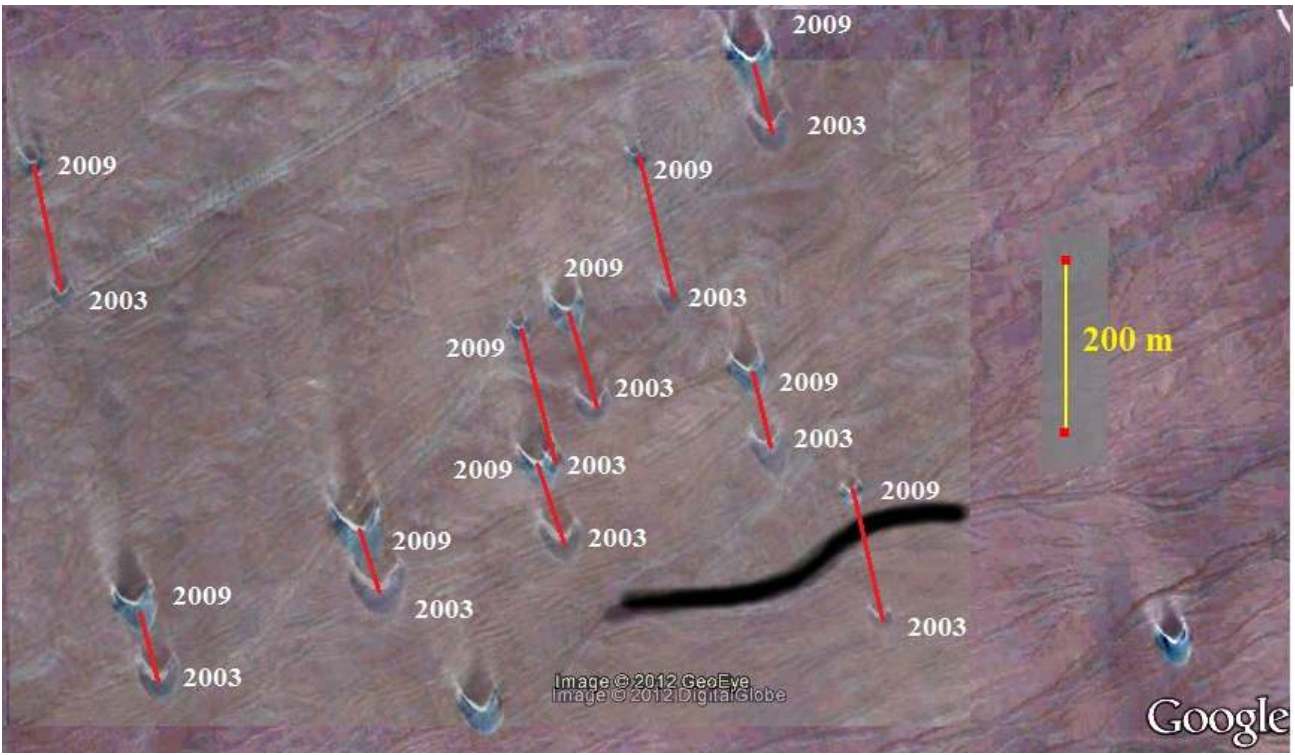

Figure 6 – Combining the images of 2003 and 2009 we can evaluate the speed of each dune.

In this period of 6.4 years, the small dunes moved of about 180 meters. The larger dune moved of about 70 meters.

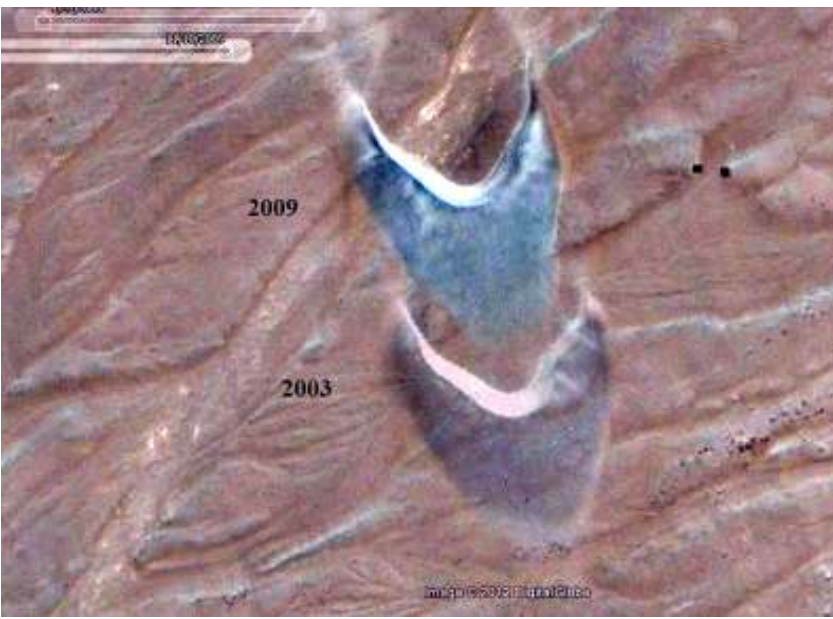
Fig.7 A dune at different times.

### 3. Analysis of a single dune and movies

In the field of the dunes of Figure 6, we can isolate one of the dune and analyze how its shape is changing during several years, for instance by the superposition of two images recorded at different times (see Figure 7). When we have several images in the time series we can even prepare a movie of the dune moving on the land. In fact, since, usually, there is a slight shift of the coordinates in the images of the series, one must be careful in preparing the movie, choosing the same reference point for all the images and adjust them correspondingly. For instance, in the Figure 7, I added two small black dots on the images to have the correct superposition.

An example of a movie is than given at http://staff.polito.it/amelia.sparavigna/DUNE/ .

To have a qualitative analysis of a single dune, or of a group of dunes, the use of a grid on the images could be interesting. For instance, if we would like to prepare a plot of the migration rate, that is the speed of each dunes with respect to its size, it would be necessary to measure distances and sizes. The distance travelled by the dune can be easily evaluated with the scales provided by Google. Let us define the "size" as the surface covered by the base of the dune. We can measure it using the grid of GIMP, as in the following example. In the Figure 8 we see two moving dunes. One is smaller and then moves faster. The figure is obtained combining three satellite images.

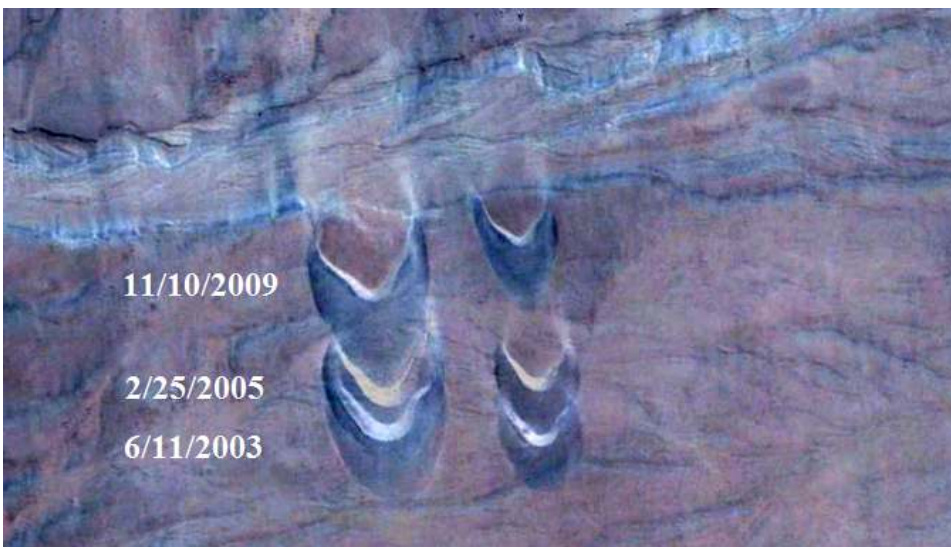Figure 8

For the size, defined as previously like the surface occupied by the dune base, we can have a rough estimate using a grid, for instance like in the Figure 9, and counting the cells occupied by the dunes.

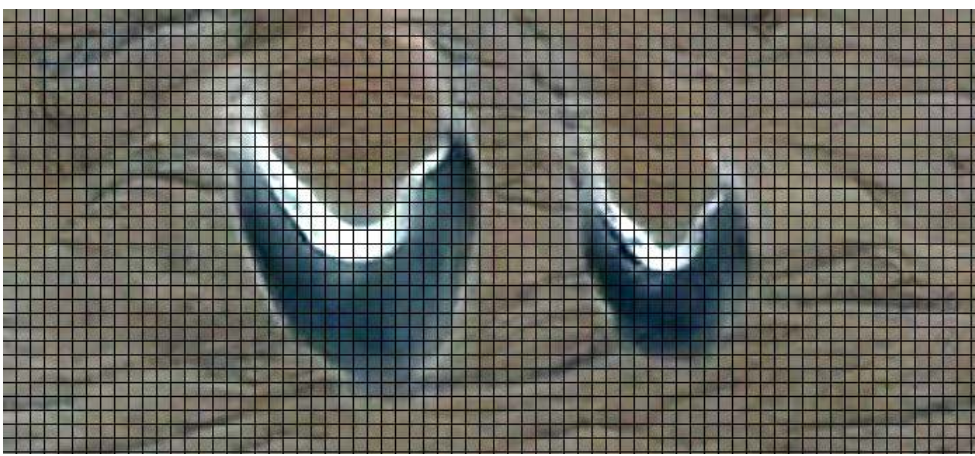Figure 9

Of course, this is a simple approach that can be improved by some image processing method, based on thresholding for instance.

   4. **Dune interaction**

Field evidence shows that two dunes can pass through one another, preserving their shape, having therefore a solitonic behavior [20]. Then the phenomenon of the dune interaction is quite rich and deserves the study of all available data. The collision of the dunes can be easily studied is the Google Earth too. Here in the following Figure 10 and 11 there are two examples of interactions, where we can see a collision, a smaller dune passing through a large one, and a fragmentation.

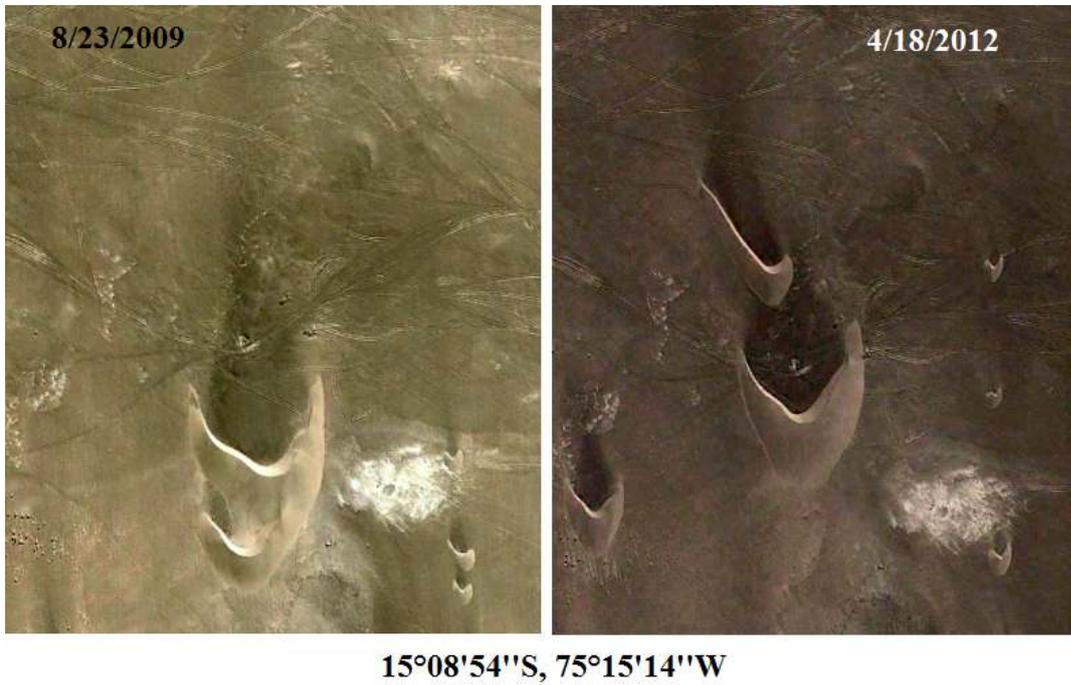

Fig.10

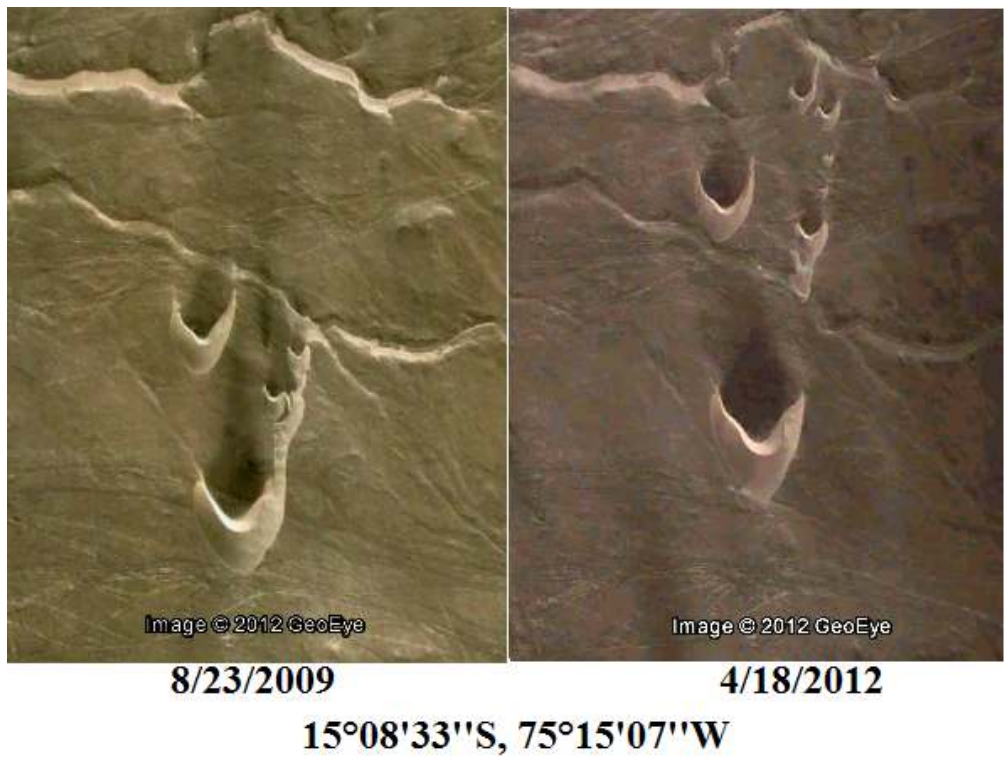

Fig.11

In this paper we have started a discussion on how some data can obtained from the time series of Google Earth. It is clear that the speed of the dunes can be easily measured. Moreover, we can find several examples of the interaction of dunes. In particular the field of the dunes of Figure 10 and 11 is under investigation to find other examples of collisions.

**References**

1. Sand dune fixation techniques, http://www.fao.org/docrep/012/i1488e/i1488e04.pdf
2. Tong-Hui Zhang, Ha-Lin Zhao, Sheng-Gong Li, Feng-Rui Li, Yasuhito Shirato, Toshiya Ohkuro, Ichiro Taniyama, A comparison of different measures for stabilizing moving sand dunes in the Horqin Sandy Land of Inner Mongolia, China, Journal of Arid Environments, Volume 58, Issue 2, July 2004, Pages 203–214
3. Guo Yu Qiu, In-Bok Lee, Hideyuki Shimizu, Yong Gao, Guodong Ding, Principles of sand dune fixation with straw checkerboard technology and its effects on the environment, Journal of Arid Environments, Volume 56, Issue 3, February 2004, Pages 449–464
4. Han Zhiwen, Wang Tao, Sun Qingwei, Dong Zhibao, Wang Xunming, Sand harm in Taklimakan Desert highway and sand control, Journal of Geographical Sciences, 13,1,(2003) 45-53.
5. R.S. Anderson, P.K. Haff, Simulation of Aeolian saltation, Science, Vol.241, 1988, 820-823
6. 3. M. Sørensen, On the rate of Aeolian sand transport, Proceedings of ICAR5/GCTESEN Joint Conference, Int. Center fro arid and semiarid lands studies, Texas Tech University, USA
7. Du He-qiang, Han Zhi-wen, Wang Tao, Sun Jia-huan, Variation of Wind Profile and Sand Flow Structure above Barchan Dune, Journal of Desert Research, 2012-01.
8. H.J. Herrmann, Aeolian transport and dune formation, published in Modelling Critical and Catastrophic Phenomena in Geoscience: A Statistical Physics Approach, eds. B.K. Chakrabarty and P. Bhattacharyya, Lecture Notes in Physics 705, Springer, 2006, 363-386.
9. The shape of dunes, H.J. Herrmann, G. Sauermann, Physica A 283 (2000) 24-30
10. H.J. Herrmann, G. Sauermann, V. Schwämmle, The morphology of dunes, Physica A, Vol.358, 2005, 30–38
11. G. Sauermann, P. Rognon, A. Poliakov, H.J. Herrmann, The shape of the barchans dunes of Southern Morocco, Geomorphology, Vol.36, 2000, 47–62
12. G. Sauermann, J.S. Andrade Jr., L.P. Maia, U.M.S. Costa, A.D. Araújo, H.J. Herrmann, Wind velocity and sand transport on a barchan dune, Geomorphology, Vol.54, 2003, 245–255
13. R.A. Bagnold, The physics of blown sand and desert dunes, London, Methuen, 1941
14. A Wind Tunnel Experiment of Aeolian Sand Transport over Wetted Coastal Sand Surface, Han Qing-jie, Qu Jian-jun, Liao Kong-tai, Zhu Shu-juan, Dong Zhi-bao, Zhang Ke-cun, ZU Rui-ping, Journal of Desert Research, 2012, Vol. 32 Issue (6) :1512-1521
15. M. Hiza Redsteer, R.C. Bogle, J.M. Vogel, Monitoring and Analysis of Sand Dune Movement and Growth on the Navajo Nation, Southwestern United States, U.S. Geological Survey, Fact Sheet 2011-3085, July 2011
16. Ding Lian-gang, Yan Ping, Du Jian-hui, Ma Yu-feng, Zhou Jing-sheng, Monitoring the state of erosion and deposition in straw checkerboard barriers based on 3D laser scanning technique, Science of Surveying and Mapping, 2009-02.
17. A.C. Sparavigna, Moving sand dunes, arXiv, 2011, Geophysics, arXiv:1112.5572, at http://arxiv.org/abs/1112.5572
18. http://www.gimp.org/
19. S. Parker Gay Jr., Observations regarding the movement of barchan sand dunes in the Nazca to Tanaca area of southern Peru, Geomorphology, Vol.27, 1999, 279-293
20. Barbara Horvat, Barchan Dunes, Seminar 2, http://www-f1.ijs.si/~rudi/sola/Sem4.pdf
21. http://en.wikipedia.org/wiki/Barchan, on 3 January 2013.